\documentclass[12pt]{article}
\usepackage[centertags]{amsmath}
\usepackage{amsfonts}
\usepackage{amssymb}
\usepackage{amsthm}
\usepackage{newlfont}

\numberwithin{equation}{section}

\newtheorem{theorem}{Theorem}
\newtheorem{lemma}{Lemma}

\begin{document}
\title{The non-Hermitian operators on the Baker-Hausdorff formula}
\author{Jack Whongius\thanks{School of Mathematical Sciences,~ Xiamen
University,~ 361005,  China.  Email:  fmsswangius@stu.xmu.edu.cn.}}

\maketitle

\par
\begin{abstract}
This paper provides a connection to the non-Hermitian operators associated with the geometric potential function $s$ and Baker-Hausdorff formula. The geometric quantum potential is considered in a precise condition.
The Ri-operator as a non-Hermitian Hamiltonian to describe the generalized quantum harmonic oscillator can be re-expressed as a more compact quantum formula by using the Baker-Hausdorff formula, more deeply, we prove some results based on the application of this formula.
\end{abstract}

\tableofcontents


\section{Introduction}

\subsection{Baker-Hausdorff formula}
Baker Hausdorff formula [1,2,3] is a very useful formula to be used in many fields. In mathematics, it can be used to give a relatively simple proof of the deep results corresponding to Lie groups and Lie algebras. In quantum mechanics, it can be realized the transformation of systematic Hamiltonian in Schrodinger's picture and Heisenberg's picture, and has many applications in perturbation theory.  Many similar transformations, such as the translation of time and space, are often realized by power exponent. In these cases, Baker-Hausdorff formula can be used in the first order.

\begin{lemma}{[1,2,3]}

The Baker-Hausdorff formula is given by
\[{{e}^{\hat{L}}}\hat{g}{{e}^{-\hat{L}}}=\hat{g}+{{\left[ \hat{L},\hat{g} \right]}_{QPB}}+\frac{1}{2!}{{\left[ \hat{L},{{\left[ \hat{L},\hat{g} \right]}_{QPB}} \right]}_{QPB}}+\frac{1}{3!}{{\left[ \hat{L},{{\left[ \hat{L},{{\left[ \hat{L},\hat{g} \right]}_{QPB}} \right]}_{QPB}} \right]}_{QPB}}+\cdots \]

\begin{proof}
  Let $f\left( y \right)={{e}^{y\hat{L}}}\hat{g}{{e}^{-y\hat{L}}}$ be given, where here $y$ is an auxiliary parameter, then
\begin{align}
  & \frac{df\left( y \right)}{dy}={{e}^{y\hat{L}}}{{\left[ \hat{L},\hat{g} \right]}_{QPB}}{{e}^{-y\hat{L}}}, \notag\\
 & \frac{{{d}^{2}}f\left( y \right)}{d{{y}^{2}}}={{e}^{y\hat{L}}}{{\left[ \hat{L},{{\left[ \hat{L},\hat{g} \right]}_{QPB}} \right]}_{QPB}}{{e}^{-y\hat{L}}}, \notag\\
 & \cdots  \notag
\end{align}
and in another way, by Taylor's expansion at $y=0$, it gets \[f\left( y \right)=f\left( 0 \right)+f'\left( 0 \right)y+\frac{1}{2!}f''\left( 0 \right){{y}^{2}}+\cdots. \]Hence, it leads to
\begin{align}
  & f\left( y \right)={{e}^{y\hat{L}}}\hat{g}{{e}^{-y\hat{L}}}=\hat{g}+\frac{1}{1!}{{\left[ \hat{L},\hat{g} \right]}_{QPB}}y+\frac{1}{2!}{{\left[ \hat{L},{{\left[ \hat{L},\hat{g} \right]}_{QPB}} \right]}_{QPB}}{{y}^{2}},  \notag\\
 & \begin{matrix}
   {} & {} & {} & {} & {} & {} & {}& {} & {} & {} \\
\end{matrix}+\frac{1}{3!}{{\left[ \hat{L},{{\left[ \hat{L},{{\left[ \hat{L},\hat{g} \right]}_{QPB}} \right]}_{QPB}} \right]}_{QPB}}{{y}^{3}}+\cdots , \notag
\end{align}Hence, $$f\left( 0 \right)=\hat{g},$$ $$f'\left( 0 \right)={{\left[ \hat{L},\hat{g} \right]}_{QPB}},$$ $$f''\left( 0 \right)={{\left[ \hat{L},{{\left[ \hat{L},\hat{g} \right]}_{QPB}} \right]}_{QPB}}.$$
Therefore, let $y=1$ be given for the derivation
\begin{align}
  & f\left( 1 \right)={{e}^{{\hat{L}}}}\hat{g}{{e}^{-\hat{L}}}=\hat{g}+\frac{1}{1!}{{\left[ \hat{L},\hat{g} \right]}_{QPB}}+\frac{1}{2!}{{\left[ \hat{L},{{\left[ \hat{L},\hat{g} \right]}_{QPB}} \right]}_{QPB}}, \notag\\
 & \begin{matrix}
   {} & {} & {} & {}  \\
\end{matrix}+\frac{1}{3!}{{\left[ \hat{L},{{\left[ \hat{L},{{\left[ \hat{L},\hat{g} \right]}_{QPB}} \right]}_{QPB}} \right]}_{QPB}}+\cdots.  \notag
\end{align}
Then we finish the proof.

\end{proof}

\end{lemma}
Note that the process of proof of the Baker-Hausdorff formula gives us more information to deeply understand the transformation of systematic Hamiltonian.
Let's set $\Lambda \left( \hat{L},\hat{g} \right)={{e}^{{\hat{L}}}}\hat{g}{{e}^{-\hat{L}}}$ for a simple consideration, then $f\left( 1 \right)=\Lambda \left( \hat{L},\hat{g} \right)$,
note that $\hat{L},\hat{g} $ are operators, for instance,
$\Lambda \left( x,\hat{g} \right)={{e}^{{x}}}\hat{g}{{e}^{-x}}$ on the position operator, $\Lambda \left( \hat{H},\hat{g} \right)={{e}^{{\hat{H}}}}\hat{g}{{e}^{-\hat{H}}}$ stands on the Hamiltonian operator. On the other side, $\Lambda \left( -\hat{L},\hat{g} \right)={{e}^{{-\hat{L}}}}\hat{g}{{e}^{\hat{L}}}$ is also to be seen for the $-\hat{L}$.

For two operators $C$ and $B$,  Baker-Hausdorff formula can be simply rewritten as
\begin{align}
  &\Lambda \left( C,B \right)= {{e}^{C}}B{{e}^{-C}}=B+{{\left[ C,B \right]}_{QPB}}+\frac{1}{2}{{\left[ C,{{\left[ C,B \right]}_{QPB}} \right]}_{QPB}}  \notag\\
 & \begin{matrix}
   {} & {} & {} & {} & {} & {}  \\
\end{matrix}+\cdots +\frac{1}{n!}{{\left[ \underbrace{C,\left[ C,\cdots  \right.\left[ C \right.}_{n~C's},{{\left. {{\left. B \right]}_{QPB}} \right]}_{QPB}} \right]}_{QPB}}+\cdots  \notag
\end{align}
Here we omit the hat over the $C,B$, therefore, it has some special results given by the following
\begin{align}
 & \Lambda \left( -C,B \right)={{e}^{-C}}B{{e}^{C}} \notag\\
 & \Lambda \left( 0,B \right)=\Lambda \left( 1,B \right)=B \notag\\
 & \Lambda \left( C,x \right)=x \notag\\
 & \Lambda \left( C,s \right)=s \notag
\end{align}
where $s$ is the geometric potential function while $x$ is the position operator.

\subsection{The G-dynamics}
The one-dimensional G-dynamics of type I is a quantum operator expressed as
\begin{equation}\label{a1}
  {{{\hat{w}}}^{\left( cl \right)}}={{b}_{c}}\left( 2u\frac{d}{dx}+u_{x}\right),
\end{equation}
and type III is ${{w}^{\left( s \right)}}=2{{b}_{c}}{{u}^{2}}$, their summation produces
 the type II ${{{\hat{w}}}^{\left( ri \right)}}={{\hat{w}}^{\left( cl \right)}}+{{w}^{\left( s \right)}}$, where line curvature $u=ds/dx$ and $u_{x}=du/dx$ have been used, $s$ is the geometric
potential function only given by the manifold space itself, and it exists as a peculia operator such as the position operator, and $b_{c}=-i\hbar/(2m)$. Due to G-dynamics of type I is a self-adjoint operator, then it helps to construct four different kinds of the non-Hermitian Hamiltonian operators: Ri-operator ${{\hat{H}}^{\left( ri \right)}}$, motor operator ${{\hat{E}}^{\left( w \right)}}$, geometric Hamiltonian operator ${{\hat{H}}^{\left( s \right)}}$ and the G-operator ${{{\hat{H}}}^{\left( gr \right)}}$ in terms of the G-dynamics of type II all as [5] shows. In particular, the motor operator and geometric Hamiltonian operator linked to the quantum potential somehow are written as

(1)~ ${{\hat{E}}^{\left( w \right)}}={{c}_{1}}{d}^{2}/d{{x}^{2}}-i\hbar {{\hat{w}}^{\left( cl \right)}}$.

(2)~ ${{\hat{H}}^{\left( s \right)}}={{c}_{1}}{{u^{2}}}-i\hbar {{\hat{w}}^{\left( cl \right)}}$.

where
 ${{c}_{1}}=-i\hbar b_{c}=-{{\hbar }^{2}}/\left( 2m \right)$.
For the basic facts we recall here, we refer to any textbook on partial differential equations. The G-dynamics of type I \eqref{a1} is a self-adjoint operator, so there exists a sequence of positive eigenvalues (going to $+\infty$) and a sequence of corresponding eigenfunctions [5].

\subsection{Geometric quantum potential and the potenergy}
The Schr\"{o}dinger equation
\begin{equation}\label{a6}
 i\hbar {\frac  {\partial \psi}{\partial t}}={{\hat{H}}^{\left( cl \right)}}\psi=\left({{c}_{1}}\nabla ^{2}+V\right)\psi
\end{equation}
is rewritten using the polar form for the wave function ${\displaystyle \psi ={{A}}e^{iS/\hbar}}$ with real-valued functions ${{A}}$ and $S$, where ${{A}}$ is the amplitude of the wave function $\psi$, and $S/\hbar$  its phase.
Quantum potential $${{Q}^{\left( cl \right)}}={{c}_{1}}\frac{{{\nabla }^{2}}{{A}}}{{{A}}}$$ as part of the Schr\"{o}dinger equation is widely studied.
The Schr\"{o}dinger equation can be also rewritten in a form of time evolution 
\[\frac{\partial \psi }{\partial t}=\frac{{{{\hat{H}}}^{\left( cl \right)}}}{i\hbar }\psi ={{\hat{f}}^{\left( cl \right)}}\psi \]
where ${{\hat{f}}^{\left( cl \right)}}=\frac{{{{\hat{H}}}^{\left( cl \right)}}}{i\hbar }=-i\frac{{{{\hat{H}}}^{\left( cl \right)}}}{\hbar }$, and then \[{{\hat{H}}^{\left( cl \right)}}=i\hbar {{\hat{f}}^{\left( cl \right)}}\]Obviously, the operator ${{\hat{f}}^{\left( cl \right)}}$ is an anti-Hermitian operator expressed as
\[{{\hat{f}}^{\left( cl \right)}}=-{{b}_{c}}{{\nabla }^{2}}-i\frac{V}{\hbar }\]

Since there are more clues to give us a hint that the quantum potential should be developed as the discovery of the G-dynamics, on this vision, it natually leads to
the general geometric quantum potential associated with the geometric potential function\footnote{geometric quantum potential ${{Q}^{\left( s \right)}}$, geometric potential function $s$, geometric potential ${{Q}^{\left( g \right)}}$ should be distinguished in a little difference}
$${{Q}^{\left( s \right)}}={{Q}^{\left( cl \right)}}+{{Q}^{\left(g \right)}}$$
where geometric potential is defined as \[{{Q}^{\left( g \right)}} ={{c}_{1}}{{\left| \nabla s \right|}^{2}}-\frac{i\hbar{{\hat{w}}^{\left( cl \right)}}{{A}}}{{{A}}}=\frac{{{\hat{H}}^{\left( s \right)}}{A}}{{A}}\]Obviously, this  geometric potential comes from the geometric Hamiltonian operator with the eigenvalues equation ${{\hat{H}}^{\left( s \right)}}{A}={{Q}^{\left( g \right)}}{A} $.
The geometric quantum potential is also expressed as
\begin{align}
 {{Q}^{\left( s \right)}} & ={{c}_{1}}\frac{\Delta {{A}}}{{{A}}}-\frac{i\hbar {{\hat{w}}^{\left( cl \right)}}{{A}}}{{{A}}}+{{c}_{1}}{{\left| \nabla s \right|}^{2}} =\frac{{{\hat{E}}^{\left( w \right)}}{{A}}}{{{A}}}+{{c}_{1}}{{\left| \nabla s \right|}^{2}} \notag
\end{align}by using the motor operator in terms of the G-dynamics of type I, and $$\frac{{{\hat{E}}^{\left( w \right)}}{{A}}}{{{A}}}=\frac{{{c}_{1}}\Delta {{A}}}{{{A}}}-\frac{i\hbar {{\hat{w}}^{\left( cl \right)}}{{A}}}{{{A}}}$$  where $\Delta={{\nabla }^{2}}$ is the Laplacian, and multi-dimensional G-dynamics of type I is $${{\hat{w}}^{\left( cl \right)}}={{b}_{c}}\left( 2u\cdot\nabla +\nabla u \right), $$ and here $u=\nabla s$ is the gradient of geometric
potential function $s$. The corresponding geometric wave equation  follows
\[i\hbar {{\hat{w}}^{\left( cl \right)}}\psi ={{\hat{H}}^{\left( \operatorname{clm} \right)}}\psi =-{{c}_{1}}\left( 2u\cdot{\nabla{\psi }}+\psi {\nabla{u}} \right),\]  In another way, using geometric Hamiltonian operator, the geometric quantum potential is then rewritten as
$${{Q}^{\left( s \right)}} ={{c}_{1}}\frac{\Delta {{A}}}{{{A}}}+\frac{{{\hat{H}}^{\left( s \right)}}{A}}{{A}}$$As it expresses, the geometric Hamiltonian operator as a supplement has revised the classical quantum potential.
As a result of the geometric potential, the potenergy ${{E}^{\left( wg \right)}}$ is given by
\[{{E}^{\left( wg \right)}}=- \frac{{i\hbar{\hat{w}}^{\left( cl \right)}}{{A}}}{{{A}}}=-i\hbar \frac{{{{\hat{w}}}^{\left( cl \right)}}A}{A}=-i\hbar {{b}_{c}}\left( 2u\cdot \nabla \ln A+\nabla u \right)\]where \[\frac{{{{\hat{w}}}^{\left( cl \right)}}A}{A}={{b}_{c}}\left( 2u\cdot \nabla \ln A+\nabla u \right)\] It appears that the potenergy holds a similar format as the quantum potential shows, actually, it can be seen this feature from the formula of the motor operator.
Accordingly, the geometric quantum potential in one-dimensional case is rewritten in a form given by
\[{{Q}^{\left( s \right)}}=\frac{{{\hat{E}}^{\left( w \right)}}{{A}}}{{A}}+{{c}_{1}}{{u}^{2}}\]
Note that quantum potential, geometric quantum potential, geometric potential are directly associated with the amplitude ${A}$ of the wave function, it implies that the precise form of the geometric quantum potential is completely determined by the expression of the amplitude.

The Schr\"{o}dinger equation and G-dynamics of type I are listed as follows for a comparison
\begin{align}
  & {{{\hat{H}}}^{\left( cl \right)}}=i\hbar {{\hat{f}}^{\left( cl \right)}},~~{{\hat{f}}^{\left( cl \right)}}=-{{b}_{c}}{{\nabla }^{2}}-i\frac{V}{\hbar } \notag\\
 & {{{\hat{H}}}^{\left( \operatorname{clm} \right)}}=i\hbar {{{\hat{w}}}^{\left( cl \right)}},~~{{{\hat{w}}}^{\left( cl \right)}}={{b}_{c}}\left( 2u\cdot \nabla +\nabla u \right)=2{{b}_{c}}u\cdot \nabla +{{b}_{c}}\nabla u \notag
\end{align}
Actually, we can put ${{\hat{f}}^{\left( cl \right)}},{{{\hat{w}}}^{\left( cl \right)}}$ together to analyze, 
\begin{align}
  & {{\hat{f}}^{\left( cl \right)}}=-{{b}_{c}}{{\nabla }^{2}}-\frac{i}{\hbar }V \notag\\
 & {{{\hat{w}}}^{\left( cl \right)}}=2{{b}_{c}}u\cdot \nabla -\frac{i}{\hbar }{{V}^{\left( r \right)}} \notag
\end{align}
where ${{V}^{\left( r \right)}}=\frac{{{\hbar }^{2}}}{2m}\nabla u$ can be treated as a kind of potential energy associated with the function $u$, their difference is that the former one is an anti-Hermitian operator while the latter one is the Hermitian operator.

\subsection{Generalized quantum harmonic oscillator }
The one-dimensional generalized quantum harmonic oscillator is described by the Ri-operator precisely written as [4]
 \begin{equation}\label{a2}
 \hat{H}^{\left( ri \right)}={{\hat{H}}}^{{\left(cl \right)}}-\frac{{{E}^{\left( s \right)}}}{2}-i\hbar {{\hat{w}}^{\left( cl \right)}}.
\end{equation}

The classical Hamiltonian operator: $\hat{H}^{\left( cl \right)}={{\hat{T}}^{\left( cl \right)}}+V\left( x \right)={{c}_{1}}\frac{{{d}^{2}}}{d{{x}^{2}}}+V\left( x \right)$.

The potential energy: $V\left( x \right)=\frac{1}{2}m{{\omega }^{2}}{{x}^{2}}$.

The imaginary geomenergy: ${{\hat{H}}^{\left( \operatorname{clm} \right)}}={{E}^{\left( \operatorname{Im} \right)}}\left( {{\hat{w}}^{\left( cl \right)}} \right)=i\hbar {{\hat{w}}^{\left( cl \right)}}$.

The one-dimensional classical kinetic energy operator: ${{\hat{T}}^{\left( cl \right)}}={{c}_{1}}\frac{{{d}^{2}}}{d{{x}^{2}}}$.

The geometric potential energy function: ${{E}^{\left( s \right)}}=-2{{c}_{1}}{{u}^{2}}$, the geometric potential energy function ${{E}^{\left( s \right)}}(u)=-2{{c}_{1}}{{u^{2}}}=i\hbar{{w}^{\left( s \right)}}$ is associated with the line curvature $u=u(x)$.\\
Based on Ri-operator, the geometric number operator can be given by
$${{\hat{N}}^{\left( r \right)}}=\hat{N}+{{\hat{N}}^{\left( s \right)}}={{a}^{\dagger }}a+{{a}^{\dagger }}b-ba-{{b}^{2}},$$
where
\begin{align}
  & {{\hat{N}}^{\left( s \right)}}={{a}^{\dagger }}b-ba-{{b}^{2}}=-i{{\hat{w}}^{\left( cl \right)}}/\omega-i{{w}^{\left( s \right)}}/2\omega,\notag\\
 & \hat{N}={{a}^{\dagger }}a=\frac{m\omega }{2\hbar }{{x}^{2}}-\frac{\hbar }{2m\omega }\frac{{{d}^{2}}}{d{{x}^{2}}}-1/2 ,\notag
\end{align}where ${b}^{2}=i{{w}^{\left( s \right)}}/2\omega=\frac{\hbar }{2m\omega }{{u}^{2}} $, $a$ is the annihilation operator while ${{a}^{\dagger }}$ is the creation operator,
and mixed term is expressed as
 $$ {{\hat{N}}^{\left( mix \right)}}(a,{a}^{\dagger },b)={{a}^{\dagger }}b-ba=-i{{\hat{w}}^{\left( cl \right)}}/\omega.$$Therefore,
the Ri-operator \eqref{a2} can be simply expressed as
$${{\hat{H}}^{\left( ri \right)}}=\hbar \omega \left( {{\hat{N}}^{\left( r \right)}}+1/2 \right)=\hbar \omega \left({{a}^{\dagger }}a+{{a}^{\dagger }}b-ba-{{b}^{2}}+1/2 \right)$$
With the help of formulae
\[  {{\hat{N}}^{\left( cg \right)}} =-{{\left[ s,\hat{N} \right]}_{QPB}}=-\frac{{{{i\hat{w}}}^{\left( cl \right)}}}{\omega },\]

\[{{\left[ s,{{{\hat{N}}}^{\left( cg \right)}} \right]}_{QPB}}=i{{w}^{\left( s \right)}}/\omega .\]Then it gets
${{\hat{N}}^{\left( s \right)}}={{\hat{N}}^{\left( cg \right)}}-{{b}^{2}}={{a}^{\dagger }}b-ba-{{b}^{2}}$, and ${{\hat{N}}^{\left( cg \right)}}={{a}^{\dagger }}b-ba$.

\section{The non-Hermitian operators on Baker-Hausdorff formula}

\subsection{Geometric potential function and Baker-Hausdorff formula}

Note that the geometric potential function is a special operator as position operator does, according to $f\left( y \right)={{e}^{y\hat{L}}}\hat{g}{{e}^{-y\hat{L}}}$, obviously,
\begin{align}
  & f\left( 0 \right)=\hat{g} ,\notag\\
 & f\left( 1 \right)={{e}^{{\hat{L}}}}\hat{g}{{e}^{-\hat{L}}} ,\notag\\
 & f\left( -1 \right)={{e}^{-\hat{L}}}\hat{g}{{e}^{{\hat{L}}}} .\notag
\end{align}
In particular, when we consider the third case, namely, let $y=-1$ be chosen, then it gets
\begin{align}
  & \Lambda \left( -\hat{L},\hat{g} \right)=f\left( -1 \right)={{e}^{-\hat{L}}}\hat{g}{{e}^{{\hat{L}}}}=\hat{g}-\frac{1}{1!}{{\left[ \hat{L},\hat{g} \right]}_{QPB}}+\frac{1}{2!}{{\left[ \hat{L},{{\left[ \hat{L},\hat{g} \right]}_{QPB}} \right]}_{QPB}}, \notag\\
 & \begin{matrix}
   {} & {} & {} & {} & {} & {} & {} & {} & {} & {}  \\
\end{matrix}-\frac{1}{3!}{{\left[ \hat{L},{{\left[ \hat{L},{{\left[ \hat{L},\hat{g} \right]}_{QPB}} \right]}_{QPB}} \right]}_{QPB}}+\cdots . \notag
\end{align}
Using this formula, here we take $\hat{L}=s$ into consideration, then the  Baker-Hausdorff formula becomes
\begin{align}\label{a3}
  &\Lambda \left( -s,\hat{g} \right)={{e}^{-s}}\hat{g}{{e}^{s}}=\hat{g}-{{\left[ s,\hat{g} \right]}_{QPB}}+\frac{1}{2!}{{\left[ s,{{\left[ s,\hat{g} \right]}_{QPB}} \right]}_{QPB}}, \\
 & \begin{matrix}
   {} & {} & {} & {} & {} & {} & {}& {} & {} & {} \\
\end{matrix}-\frac{1}{3!}{{\left[ s,{{\left[ s,{{\left[ s,\hat{g} \right]}_{QPB}} \right]}_{QPB}} \right]}_{QPB}}+\cdots . \notag
\end{align}Actually, this formula can be regarded as a transformation of the operator $\hat{g}$ associated with the geometric potential function, namely, $$\hat{g}\rightarrow  \Lambda \left( -s,\hat{g} \right)={{e}^{-s}}\hat{g}{{e}^{s}}.$$ If $s=Constants$ or $s=0$ holds, then $\hat{g}={{e}^{-s}}\hat{g}{{e}^{s}}$. In general, the $s$ appears as a function. As it shows, the terms ${{\left[ s,\hat{g} \right]}_{QPB}}$ plays a central role in developing the non-Hermitian operators connected to the quantum covariant Poisson bracket. If $\hat{g}=x$ is a position operator, then ${{\left[ s,x \right]}_{QPB}}=0$ is given for $x={{e}^{-s}}x{{e}^{s}}$, this clearly indicates that the geometric potential function $s$ is a basic variable similar to the position operator $x$.

We denote
\begin{equation}\label{a4}
  \lambda\left( s,\hat{g}\right)=-{{\left[ s,\hat{g} \right]}_{QPB}}+\frac{1}{2!}{{\left[ s,{{\left[ s,\hat{g} \right]}_{QPB}} \right]}_{QPB}}-\frac{1}{3!}{{\left[ s,{{\left[ s,{{\left[ s,\hat{g} \right]}_{QPB}} \right]}_{QPB}} \right]}_{QPB}}+\cdots.
\end{equation}It can be understood as some kinds of the interactions between the environment represented by $s$ and operator $\hat{g} $.
It's obvious to see that the non-trivial result $$\lambda\left( s,\hat{g}\right)={{e}^{-s}}\hat{g}{{e}^{s}}-\hat{g},$$ holds for all operator $\hat{g}$. Conversely,  $$\Lambda \left( -s,\hat{g} \right)={{e}^{-s}}\hat{g}{{e}^{s}}=\hat{g}+\lambda\left( s,\hat{g}\right),$$ can be regarded as a standard procedure for a given operator $\hat{g}$ to be generalized to directly associate with the basic parameter $s$. Therefore, some special results follow $$\lambda\left( Constants,\hat{g}\right)=\lambda\left( 0,\hat{g}\right)=\lambda\left( s,x\right)=0.$$As a result, we can conclude that if the condition
${{\left[ s,\hat{g} \right]}_{QPB}}=0$ is given, then $\lambda\left( s,\hat{g}\right)=0$ always stands for $$\Lambda \left( -s,\hat{g} \right)={{e}^{-s}}\hat{g}{{e}^{s}}=\hat{g},$$ such as the position operator, we always consider the non-constant function $s$ for ${{\left[ s,\hat{g} \right]}_{QPB}}=0$ anyway.
General speaking, $\lambda\left( s,\hat{g}\right)\neq 0$ holds for the non-trivial cases. Given a function $f$, and the Taylor's expansion of the exponential function
, then \[\lambda \left( s,\hat{g} \right)f={{e}^{-s}}\hat{g}\left( {{e}^{s}}f \right)-\hat{g}f=\left( 1-s+\frac{1}{2}{{s}^{2}}+\cdots  \right)\hat{g}\left( 1+s+\cdots  \right)f-\hat{g}f.\]
According to the \eqref{a4}, if ${{\left[ s,\hat{g} \right]}_{QPB}}=h$ is a functional form, then \eqref{a4} directly becomes $\lambda\left( s,\hat{g}\right)=-h$, it means $${{\left[ s,{{\left[ s,\hat{g} \right]}_{QPB}} \right]}_{QPB}}={{\left[ s,h \right]}_{QPB}}=0.$$ In this time, it gets
$\Lambda \left( -s,\hat{g} \right)={{e}^{-s}}\hat{g}{{e}^{s}}=\hat{g}-h$, in the following discussions, we will see lots of this cases.

\subsection{The case of exponential function and the amplitude}

Inspired by the Baker-Hausdorff formula and the geometric potential function, the exponential function ${{e}^{s}}$ can be treated as part of the amplitude, in order to claim this point,  let's consider a case of amplitude such as ${{A}}={{C}_{0}}{{e}^{s}}$, where $s$ is a geometric potential function,  on such assumption, by a direct computation, the quantum potential can be calculated accordingly, more specifically,
\[{{Q}^{\left( cl \right)}}={{c}_{1}}\frac{{{\nabla }^{2}}{{A}}}{{{A}}}={{c}_{1}}\frac{{{C}_{0}}{{\nabla }^{2}}{{e}^{s}}}{{{A}}}={{c}_{1}}\left( {{\nabla }^{2}}s+\nabla s\cdot \nabla s \right)\]
Obviously, it's equal to the structural c-energy ${{T}^{\left( c \right)}}$, that is to say, ${{T}^{\left( c \right)}}={{Q}^{\left( cl \right)}}$ in this case. Meanwhile, it obtains \[\frac{{{\hat{w}}^{\left( cl \right)}}{{A}}}{{A}}={{b }_{c}}\left( \Delta s+2\nabla s\cdot \nabla s \right)\]It leads to a precise formulation of the potenergy ${{E}^{\left( wg \right)}}$
\begin{align}
{{E}^{\left( wg \right)}}  &=-\frac{i\hbar {{{\hat{w}}}^{\left( cl \right)}}{{A}}}{{A}}={{c}_{1}}\left( \Delta s+2\nabla s\cdot \nabla s \right) \notag\\
 & ={{Q}^{\left( cl \right)}}-{{E}^{\left( s \right)}}/2 \notag
\end{align}
Conversely, it gets  ${{Q}^{\left( cl \right)}}={{E}^{\left( s \right)}}/2+{{E}^{\left( wg \right)}}$.
It reveals the deep connection between the potenergy and the quantum potential in this case, the quantum potential as a part can be induced by the potenergy.
The geometric quantum potential is then given by
\begin{align}
 {{Q}^{\left( s \right)}} ={{Q}^{\left( cl \right)}}+{{E}^{\left( wg \right)}}+{{c}_{1}}{{\left| \nabla s \right|}^{2}}=2{{Q}^{\left( cl \right)}}-{{E}^{\left( s \right)}} \notag
\end{align}
In fact, this case tells us that the geometric quantum potential is more wider to explain some quantum problems, such as the problems caused by the environment, and it's valid to choose the proper amplitude ${A}$ of the wave function to get certain energy, various energy in a wider quantum system together have interactions to each other.

For the case of the exponential function ${{e}^{s}}$ as part of the amplitude,  let's consider a case of amplitude such as ${{A}}={{C}_{0}}{{e}^{s}}$,
in non-relativistic quantum mechanics, the wave function can be chosen as a form like
${{\psi }}=C_{0}{{e}^{{s}}}{{e}^{i\frac{S}{\hbar } }}$,  where
$s$ is geometric potential function as a real valued functions,
the gradient of this wave function is computed as
\[\nabla {{\psi }}=\left( \frac{i}{\hbar }\nabla S+u \right){{\psi }}\]
where $\nabla$ denotes gradient operator, and $u=\nabla s$.
We are thus led to the new evolution law for Bohmian mechanics.  Therefore, then
\[\frac{\nabla {{\psi }}}{{{\psi }}}=\frac{i}{\hbar }\nabla S+u \]
new evolution law for Bohmian mechanics is accordingly obtained $\frac{dQ}{dt}=a_{c}\frac{\nabla {{\psi }}}{{{\psi }}}$, precisely, the more specifical details follows
\[\frac{dQ}{dt}=\frac{i}{m }\nabla S+a_{c}u\]where $m$ is the mass of particles, and $a_{c}=\hbar/m$.
In this case for the Bohmian mechanics, we do not take the imaginary part only, as it can be seen, the real part actually exists which is a real velocity induced by the  geometric potential function,
\begin{align}
  &{{v}^{\left( e \right)}}= \operatorname{Re}\frac{dQ}{dt}=\frac{\hbar }{m}u=a_{c}u\notag\\
 & {{v}^{\left( m \right)}}=\operatorname{Im}\frac{dQ}{dt}=\frac{ \nabla S }{m} \notag
\end{align}Obviously, it has certain form $\frac{d{{Q}}}{dt}={{v}^{\left( e \right)}}+i{{v}^{\left( m \right)}}$.

Notice that ${{v}^{\left( e \right)}}=\frac{\hbar }{m}u$ is a real quantum velocity in which corresponding momentum ${{p}^{\left( e \right)}}=\hbar u=m{{v}^{\left( e \right)}}$ to guide how particles move along the path created by the vector field $\nabla s$ induced by the  the geometric
potential function or structural function $s$ on space manifolds.

\subsection{Geometric number operator on Baker-Hausdorff formula}
We apply this formula directly associated with the geometric potential function $s$ to the one-dimensional generalized quantum harmonic oscillator.
Then it leads to the theorem.
\begin{theorem}
The geometric number operator can be given by
\[{{{\hat{N}}}^{\left( r \right)}} ={{e}^{-s}}\hat{N}{{e}^{s}}=\hat{N}-{{\left[ s,\hat{N} \right]}_{QPB}}+\frac{1}{2}{{\left[ s,{{\left[ s,\hat{N} \right]}_{QPB}} \right]}_{QPB}}. \]
 \begin{proof}
By direct evaluation, it has
\begin{align}
 {{e}^{-s}}\hat{N}\left( {{e}^{s}}f \right) & =\frac{m\omega }{2\hbar }{{x}^{2}}f-\frac{\hbar }{2m\omega }{{e}^{-s}}\frac{{{d}^{2}}\left( {{e}^{s}}f \right)}{d{{x}^{2}}}-\frac{1}{2}f ,\notag\\
 & =\frac{m\omega }{2\hbar }{{x}^{2}}f-\frac{\hbar }{2m\omega }{{f}_{xx}}-\frac{1}{2}f-\frac{\hbar }{2m\omega }\left( 2u{{f}_{x}}+{{u}_{x}}f \right)-\frac{\hbar }{2m\omega }{{u}^{2}}f, \notag\\
 & =\hat{N}f-\frac{\hbar }{2m\omega }\left( 2u{{f}_{x}}+{{u}_{x}}f \right)-\frac{\hbar }{2m\omega }{{u}^{2}}f \notag\\
 & =\hat{N}f-\frac{i{{{\hat{w}}}^{\left( cl \right)}}f}{\omega }-\frac{i{{w}^{\left( s \right)}}}{2\omega }f ,\notag\\
 & =\hat{N}f+{{{\hat{N}}}^{\left( s \right)}}f \notag\\
 & ={{{\hat{N}}}^{\left( r \right)}}f, \notag
\end{align}
where \[{{e}^{-s}}\frac{{{d}^{2}}\left( {{e}^{s}}f \right)}{d{{x}^{2}}}={{f}_{xx}}+2u{{f}_{x}}+{{u}_{x}}f+{{u}^{2}}f,\]
and
\begin{align}
  & {{{\hat{N}}}^{\left( cg \right)}}=-{{\left[ s,\hat{N} \right]}_{QPB}}=-\frac{i{{{\hat{w}}}^{\left( cl \right)}}}{\omega } ,\notag\\
 & {{\left[ s,{{\left[ s,\hat{N} \right]}_{QPB}} \right]}_{QPB}}=-i{{w}^{\left( s \right)}}/\omega.  \notag
\end{align}
By the way, it always has
$-\frac{1}{3!}{{\left[ s,{{\left[ s,{{\left[ s,\hat{N} \right]}_{QPB}} \right]}_{QPB}} \right]}_{QPB}}=0$ and the terms of the higher order are equal to zero.
Therefore, we complete the proof.

\end{proof}

\end{theorem}

It is led to another form of the geometric number operator, it shows
\begin{align}
{{{\hat{N}}}^{\left( r \right)}}  &=\hat{N}-{{\left[ s,\hat{N} \right]}_{QPB}}+\frac{1}{2}\left[ s,{{\left[ s,\hat{N} \right]}_{QPB}} \right] =\hat{N}-\frac{i{{{\hat{w}}}^{\left( cl \right)}}}{\omega }-\frac{i{{w}^{\left( s \right)}}}{2\omega }. \notag
\end{align}
Note that the formula of the geometric number operator is given by the combinations of the elements ${{a}^{\dagger }},a,b$ here that amazingly corresponds to the Baker-Hausdorff formula in terms of exponential function of the geometric potential function $s$, this self-consistency implies that the geometric number operator is valid to be built for the extension of the quantum mechanics. At the same time, it also states that the geometric potential function $s$ is a fundamental parameter in quantum mechanics that can't be ignored, just like the position operator.

When Baker-Hausdorff formula comes to the geometric annihilation operator and geometric creation operators that are written as  $${{a}^{\left( s \right)}}=a-{{\left[ s,a \right]}_{QPB}},~~~~{{a}^{\left( s \right)}}^{\dagger }={{a}^{\dagger }}-{{\left[ s,{{a}^{\dagger }} \right]}_{QPB}},$$
where $a,{{a}^{\dagger }}$ are annihilation operator and creation operators respectively. Due to the facts $${{\left[ s,{{a}^{\dagger }} \right]}_{QPB}}=b,~~~{{\left[ s,a \right]}_{QPB}}=-b,$$ where $b=\sqrt{\frac{\hbar }{2m\omega }}u$, then it indicates that all the high order terms vanish, that is to say,
$${{\left[ s,{{\left[ s,a \right]}_{QPB}} \right]}_{QPB}}={{\left[ s,{{\left[ s,{{a}^{\dagger }} \right]}_{QPB}} \right]}_{QPB}}=0.$$
As a result,  the geometric annihilation operator and geometric creation operators can be rewritten as
\[{{a}^{\left( s \right)}}={{e}^{-s}}a{{e}^{s}},~~{{a}^{\left( s \right)\dagger }}={{e}^{-s}}{{a}^{\dagger }}{{e}^{s}}. \]
As a consequence, the geometric annihilation operator and geometric creation operators can be expressed as two different way, clearly, we can choose one of expressions to show what we want to deliver.

\subsection{ Ri-operator on the Baker-Hausdorff formula}
In this section, we will show how Ri-operator and geomentum operator can be rewritten in another compact form by using the Baker-Hausdorff formula.
As for the Ri-operator given by \eqref{a2}, inspired by the Baker-Hausdorff formula, it can be expressed as
\begin{equation}\label{a5}
  {{\hat{H}}^{\left( ri \right)}}={{e}^{-s}}{{\hat{H}}^{\left( cl \right)}}{{e}^{s}}={{\hat{H}}^{\left( cl \right)}}-{{\left[ s,{{\hat{H}}^{\left( cl \right)}} \right]}_{QPB}}+\frac{1}{2}{{\left[ s,{{\left[ s,{{\hat{H}}^{\left( cl \right)}} \right]}_{QPB}} \right]}_{QPB}},
\end{equation}
where
$$i\hbar {{{\hat{w}}}^{\left( cl \right)}}={{\left[ s,{{\hat{H}}^{\left( cl\right)}}\right]}_{QPB}},$$
 $${{E}^{\left( s \right)}}=-{{\left[ s,{{\left[ s,{{{\hat{H}}}^{\left( cl \right)}} \right]}_{QPB}} \right]}_{QPB}},$$.
$${{\left[ s,{{\left[ s,{{\left[ s,{{{\hat{H}}}^{\left( cl \right)}}  \right]}_{QPB}} \right]}_{QPB}} \right]}_{QPB}}=0,$$
and the terms of the higher order vanish.
 \begin{proof}
By direct evaluation, it has
\begin{align}
 {{{\hat{H}}}^{\left( ri \right)}}f &={{e}^{-s}}{{{\hat{H}}}^{\left( cl \right)}}\left( {{e}^{s}}f \right)={{e}^{-s}}\left( {{{\hat{T}}}^{\left( cl \right)}}\left( {{e}^{s}}f \right)+{{e}^{s}}fV\left( x \right) \right) , \notag\\
 & ={{e}^{-s}}{{{\hat{T}}}^{\left( cl \right)}}\left( {{e}^{s}}f \right)+fV\left( x \right),  \notag\\
 & ={{c}_{1}}{{e}^{-s}}\frac{{{d}^{2}}}{d{{x}^{2}}}\left( {{e}^{s}}f \right)+fV\left( x \right),  \notag\\
 & ={{c}_{1}}\left( {{f}_{xx}}+2u{{f}_{x}}+{{u}_{x}}f+{{u}^{2}}f \right)+fV\left( x \right),  \notag\\
 & ={{c}_{1}}{{f}_{xx}}+fV\left( x \right)+{{c}_{1}}\left( 2u\frac{d}{dx}+{{u}_{x}} \right)f+{{c}_{1}}{{u}^{2}}f,  \notag\\
 & ={{{\hat{H}}}^{\left( cl \right)}}f+{{c}_{1}}{{u}^{2}}f+{{c}_{1}}\left( 2u\frac{d}{dx}+{{u}_{x}} \right)f,  \notag\\
 & ={{{\hat{H}}}^{\left( cl \right)}}f-\frac{{{E}^{\left( s \right)}}}{2}f-i\hbar {{{\hat{w}}}^{\left( cl \right)}}f , \notag\\
 & =\left( {{{\hat{H}}}^{\left( cl \right)}}-\frac{{{E}^{\left( s \right)}}}{2}-i\hbar {{{\hat{w}}}^{\left( cl \right)}} \right)f.  \notag
\end{align}Therefore,
\begin{align}
  & {{{\hat{H}}}^{\left( ri \right)}}={{e}^{-s}}{{\hat{H}}^{\left( cl \right)}}{{e}^{s}}={{{\hat{H}}}^{\left( cl \right)}}-\frac{{{E}^{\left( s \right)}}}{2}-i\hbar {{{\hat{w}}}^{\left( cl \right)}} ,\notag\\
 & {{E}^{\left( \operatorname{Im} \right)}}\left( {{{\hat{w}}}^{\left( cl \right)}} \right)=i\hbar {{{\hat{w}}}^{\left( cl \right)}}={{\left[ s,{{{\hat{H}}}^{\left( cl \right)}} \right]}_{QPB}}, \notag\\
 -{{E}^{\left( s \right)}}&={{\left[ s,{{\left[ s,{{{\hat{H}}}^{\left( cl \right)}} \right]}_{QPB}} \right]}_{QPB}}, \notag\\
 & =i\hbar {{\left[ s,{{{\hat{w}}}^{\left( cl \right)}} \right]}_{QPB}}\notag\\
 &=-i\hbar {{w}^{\left( s \right)}},\notag\\
 &=-2i\hbar {{b}_{c}}{{u}^{2}}=-\frac{{{\hbar }^{2}}}{m}{{u}^{2}}=2{{c}_{1}}{{u}^{2}} \notag
\end{align}and \[{{\hat{T}}^{\left( ri \right)}}={{e}^{-s}}{{\hat{T}}^{\left( cl \right)}}{{e}^{s}}={{c}_{1}}\frac{{{d}^{2}}}{d{{x}^{2}}}-\frac{{{E}^{\left( s \right)}}}{2}-i\hbar {{\hat{w}}^{\left( cl \right)}},\]where $2i\hbar {{b}_{c}}=\frac{{{\hbar }^{2}}}{m}$,
and the terms of the higher order are equal to zero.
Therefore, we complete the proof.

\end{proof}
Note that the Ri-operator given by \eqref{a5} is another equivalent formula of the formula  \eqref{a2}, and obviously, the \eqref{a5} is more simple and formal way to realize how the Ri-operator has been constructed.
Thusly, it obtains ${{e}^{s}}{{\hat{H}}^{\left( ri \right)}}={{\hat{H}}^{\left( cl \right)}}{{e}^{s}}$. According to the formula \eqref{a4}, it directly deduces
\[\lambda \left( s,{{\hat{H}}^{\left( cl \right)}} \right)=-{{\left[ s,{{\hat{H}}^{\left( cl \right)}} \right]}_{QPB}}+\frac{1}{2}{{\left[ s,{{\left[ s,{{\hat{H}}^{\left( cl \right)}} \right]}_{QPB}} \right]}_{QPB}}=-i\hbar {{\hat{w}}^{\left( cl \right)}} -\frac{{{E}^{\left( s \right)}}}{2}.\]Hence, it gets
the geometric Hamiltonian operator ${{\hat{H}}^{\left( s \right)}}=\lambda \left( s,{{\hat{H}}^{\left( cl \right)}} \right)$.
As a consequence, the Ri-operator \eqref{a2} can be rewritten in a simple form
\[{{\hat{H}}^{\left( ri \right)}}={{e}^{-s}}{{\hat{H}}^{\left( cl \right)}}{{e}^{s}}={{\hat{H}}^{\left( cl \right)}}+\lambda \left( s,{{\hat{H}}^{\left( cl \right)}} \right)={{\hat{H}}^{\left( cl \right)}}+{{\hat{H}}^{\left( s \right)}}.\]This implies the valid introduction of the geometric potential function $s$, by transformation, this gives us the essence of the covariant quantum mechanics.

Using the Baker-Hausdorff formula, it helps us better understand the Ri-operator, we can realize that the zero-order is the classical Hamiltonian operator ${{\hat{H}}^{\left( cl \right)}}$, the first-order leads to the imaginary geomenergy $i\hbar {{{\hat{w}}}^{\left( cl \right)}}={{\left[ s,{{{\hat{H}}}^{\left( cl \right)}} \right]}_{QPB}} $ induced by the G-dynamics of the type I, and second-order deduces the geometric potential energy function ${{E}^{\left( s \right)}}=-{{\left[ s,{{\left[ s,{{{\hat{H}}}^{\left( cl \right)}} \right]}_{QPB}} \right]}_{QPB}}=i\hbar {{w}^{\left( s \right)}}$ induced by the G-dynamics of the type III.

This situation also goes to the geomentum operator
\[{{\hat{p}}^{\left( ri \right)}}={{e}^{-s}}{{\hat{p}}^{\left( cl \right)}}{{e}^{s}}={{\hat{p}}^{\left( cl \right)}}-{{\left[ s,{{\hat{p}}^{\left( cl \right)}} \right]}_{QPB}}, \]
where $i\hbar u={{\left[ s,{{{\hat{p}}}^{\left( cl \right)}}\left( t \right) \right]}_{QPB}}$, and ${{\left[ s,{{\left[ s,{{\hat{p}}^{\left( cl \right)}} \right]}_{QPB}} \right]}_{QPB}}=0$ holds, the high order terms disappear. Therefore, the geomentum operator can be given in a form
\[{{\hat{p}}^{\left( ri \right)}}={{\hat{p}}^{\left( cl \right)}}+\lambda \left( s,{{\hat{p}}^{\left( cl \right)}} \right),\]
where $\lambda \left( s,{{\hat{p}}^{\left( cl \right)}} \right)=-{{\left[ s,{{\hat{p}}^{\left( cl \right)}} \right]}_{QPB}}=-i\hbar u$.

If we apply the \eqref{a3} to the one-dimensional G-dynamics of type I \eqref{a1}, it produces
\[{{\hat{w}}^{\left( ri \right)}}={{e}^{-s}}{{\hat{w}}^{\left( cl \right)}}{{e}^{s}}={{\hat{w}}^{\left( cl \right)}}-{{\left[ s,{{{\hat{w}}}^{\left( cl \right)}} \right]}_{QPB}}+\frac{1}{2}{{\left[ s,{{\left[ s,{{{\hat{w}}}^{\left( cl \right)}} \right]}_{QPB}} \right]}_{QPB}}+\cdots, \]
where the G-dynamics of type III ${{{{w}}}^{\left( s \right)}}=-{{\left[ s,{{{\hat{w}}}^{\left( cl \right)}} \right]}_{QPB}}$, and then $${{\left[ s,{{\left[ s,{{{\hat{w}}}^{\left( cl \right)}} \right]}_{QPB}} \right]}_{QPB}}=0.$$ Hence, it gets the G-dynamics of type II $${{\hat{w}}^{\left( ri \right)}}={{e}^{-s}}{{\hat{w}}^{\left( cl \right)}}{{e}^{s}}={{\hat{w}}^{\left( cl \right)}}-{{\left[ s,{{{\hat{w}}}^{\left( cl \right)}} \right]}_{QPB}}$$
Therefore, the Baker-Hausdorff formula enhances our understandings of the non-Hermitian operators directly linked to the geometric potential function $s$.

\section{Conclusions}
The Baker-Hausdorff formula is helpful to provide more explanations on the non-Hermitian operators induced by the geometric potential function $s$, it leads to the non-trivial result $\lambda\left( s,\hat{g}\right)={{e}^{-s}}\hat{g}{{e}^{s}}-\hat{g}$ holds for all operator $\hat{g}$, this is all connected to the geometric potential function $s$
 $$\Lambda \left( -s,\hat{g} \right)={{e}^{-s}}\hat{g}{{e}^{s}}=\hat{g}+\lambda\left( s,\hat{g}\right).$$
 By using this standard formula, it can widely produce the non-Hermitian operators in most of time.
In this way, it's natural to explain and simplify the new non-Hermitian operators we have discovered so far. By using the exponential function ${{e}^{s}}$, the geometric quantum potential is considered in a precise condition with the amplitude associated with the exponential function ${{e}^{s}}$. To sum up,

The Ri-operator: ${{{\hat{H}}}^{\left( ri \right)}}={{e}^{-s}}{{\hat{H}}^{\left( cl \right)}}{{e}^{s}}=\Lambda \left( -s,{{\hat{H}}^{\left( cl \right)}} \right)$,

The geometrinetic energy operator: ${{\hat{T}}^{\left( ri \right)}}={{e}^{-s}}{{\hat{T}}^{\left( cl \right)}}{{e}^{s}}=\Lambda \left( -s,{{\hat{T}}^{\left( cl \right)}} \right)$,

The type II: ${{\hat{w}}^{\left( ri \right)}}={{e}^{-s}}{{\hat{w}}^{\left( cl \right)}}{{e}^{s}}=\Lambda \left( -s,{{\hat{w}}^{\left( cl \right)}} \right)$,

The geomentum operator: ${{\hat{p}}^{\left( ri \right)}}={{e}^{-s}}{{\hat{p}}^{\left( cl \right)}}{{e}^{s}}=\Lambda \left( -s,{{\hat{p}}^{\left( cl \right)}} \right)$,

The geometric annihilation operator: ${{a}^{\left( s \right)}}={{e}^{-s}}a{{e}^{s}}=\Lambda \left( -s,a \right)$,

The geometric creation operator: ${{a}^{\left( s \right)\dagger }}={{e}^{-s}}{{a}^{\dagger }}{{e}^{s}}=\Lambda \left( -s,{a}^{\dagger }\right)$,

The geometric number operator: ${{{\hat{N}}}^{\left( r \right)}} ={{e}^{-s}}\hat{N}{{e}^{s}}=\Lambda \left( -s,\hat{N}\right)$.\\
Their common features are that the orders are not more than third-order.

\section*{References}
\ \ \

\par [1]
J.J. Sakurai. Modern Quantum Mechanics (Revised Edition)[M]. Addison Wesley. 1993.

\par [2]
Yong He. Derivation of Baker-Hausdorff formula using the operator's matrix representation [J]. College Physics, 2015,34 (1): 30-31.
\par [3]
Shixun Zhou. Course of Quantum Mechanics (Second Edition)[M]. Higher Education Press, 2008.
\par [4]
Gen Wang. Geometric quantization rules in QCPB theory. arXiv preprint,	arXiv:2005.11141

\par [5]
 J Whongius. Eigenvalues of the generalized Laplacian and the G-dynamics of type I. arXiv preprint, 	arXiv:2211.14783.

\end{document}